\documentclass[sn-basic]{sn-jnl}

\usepackage[T1]{fontenc}
\usepackage{ae,aecompl}
\usepackage{graphicx}	
\usepackage{amsmath}	
\usepackage{amssymb}	
\usepackage{subfig}
\usepackage{hyperref}
\usepackage{adjustbox}

\usepackage{lmodern}
\usepackage[T1]{fontenc}
\usepackage{ae}
\usepackage[compatibility=false]{caption}
\usepackage{color}
\usepackage{upgreek}
\usepackage{float}
\begin{document}
\title[cosmological many body problem]{The Fluctuation Theory, Critical Phenomena and Gravitational Clustering of Galaxies}

\author*[1]{\fnm{Khan} \sur{M.S.}}\email{mskhan62@rediffmail.com}


\author[2]{\fnm{ Mohamed H.} \sur{Abdullah}}


\author[1]{\fnm{ Zahir} \sur{Shah}}


\author[1]{\fnm{ Owais} \sur{Farooq}}

\author[3]{\fnm{ Khan} \sur{Azmat}}

\affil*[1]{\orgdiv{Department of Physics}, \orgname{Central University of Kashmir}, \orgaddress{\city{Ganderbal}, \postcode{1 191311},  \country{India}}}

\affil[2]{\orgdiv{Institute of Management and Information Technologies}, \orgname{Chiba University}, \postcode{263-8522}, \country{Japan}}

\affil[3]{\orgdiv{Department of Electrical Engineering}, \orgname{Jamia Millia Islamia}, \orgaddress{\city{New Delhi}, \postcode{110025},  \country{India}}}


\abstract{ We investigate the phenomenon of clustering of galaxies in an expanding universe by applying the fluctuation theory. We evaluate the fluctuation moments for the number of particles as well as the correlated fluctuations for number and energy of particles (galaxies), clustering under their mutual gravitation. The correlated fluctuations $<\Delta N\Delta U>$ show interesting results. The value of $<\Delta N>$ can be both positive as well as negative, because it is the difference between $N$ and the mean value of $N$. A negative $<\Delta N>$ corresponds to regions of under density and positive $<\Delta N>$ corresponds to regions of over density, as described by the clustering parameter $b$. The present work is concerned in the region $b\ge 0$, at which gravitational interaction has already started causing the galaxies to cluster. Thus for this work the value of $<\Delta N>$ is positive. Similarly, the energy fluctuations $<\Delta U>$ can also be both positive as well as negative. For large correlations, the overdense regions typically have negative total energy and underdense regions have usually positive total energy. The critical value at which this switch occurs has been calculated analytically. The results obtained by fluctuation theory closely match with those obtained earlier by Specific heat analysis and Lee Yang theory.
The evaluation has been extended to multicomponent systems, having a variety of masses. It has been found that the gravitational clustering of galaxies is more sensitive to mass ratios and less sensitive to number densities of galaxies. This means there is little effect of $\nu$ (number density) but significant effect of $\mu$ (mass) of galaxies on the clustering phenomenon. The clustering of galaxies is quicker when mass of individual galaxies increases. They become nuclei for condensation. As the mass of galaxies increases, the transition from positive to negative energy occurs at a higher stage of clustering as compared to a single component system.}

\keywords{large-scale structure of Universe -- cosmology: theory : galaxies: clusters: general -- analytical.}

\maketitle

\section{Introduction}\label{sec1}
Gravitaional particles (galaxies) aggregate together to form bound systems. They strive to form more and more clustered systems and never reach a final equilibrium. In gravitational systems, the time scale for the clustering arrangements may become very large as compared to the normal laboratory systems. The study of these gravitational clusters is vital to our understanding of the begining, formation and evolution of the large scale structure of the universe. Moreover, they are also important for studying the models of the structure formation as well as constraining the cosmological parameters \citep{2004ApJ...608..636L, 2005ApJ...626..795S, 2011Ap&SS.336..447M, 2011ApJ...729..123Y, 2011pchm.conf..597Y}. 
In cosmological many body problem, the gravitational clustering has many features of a phase transition but gravitaional phase transitions differ from the ordinary laboratory phase transitions in many ways. While as the ordinary chemical phase transitions change the systen abruptly, from one state to another, gravitational phase transitions are slowly evolving. The gravitational phase transitions do not occur over all the scales in a short time. In an infinite homogenous gravitating system, the nearest particles (galaxies) will cluster first, then these clusters with the surrounding unclustered particles and then the clusters with the clusters. So there is a sequence of phase transitions occuring in the system. However unlike chemical phase transitions, these gravitational phase transitions never reach a final equilibrium state, striving for more and more clustered arrangement. It is a sort of continous phase transitions occuring in the system \citep{1987gpsg.book.....S, 1980lssu.book.....P, 1993ppc..book.....P, 2012ApJ...745...87Y}.

The Gravitaional phase transitions were studied by \citet{2010ApJ...720.1246S} and \citet{2012MNRAS.421.2629K, 2013Ap&SS.348..211K} utilizing the Lee Yang theory and Specific heat analysis. In the present work we study the critical phenomena and gravitational phase transitions in single as well as multicomponent systems of particles (galaxies) on the basis of fluctuation theory. This is because the fluctuations in a system are constantly probing the possibility of a phase transition. We extend our analysis to multicomponent systems because the real universe contains particles (galaxies) having a variety of masses. In sections 2 and 3, we apply fluctuation theory to single as well as multicomponent systems respectively and analytically calculate various fluctuation moments. We examine the correlated moments and find that there is a particular value of the clustering parameter at which a transition from positive to negative total energy takes place. It is interesting to find that the results obtained from fluctuation theory match with those obtained earlier from specific heat analysis and Lee Yang theory \citep{antonov1962english, 1990PhR...188..285P}.

\section{Fluctuation Theory}\label{sec2}

The fluctuations are common in thermodynamic systems and an essential part of the thermodynamic equilibrium. It is a common phenomenon that the macroscopic quantities will fluctuate around their average values, from one region to another or within the sub-regions of the system. The fluctuations are normally small but in gravitational systems the fluctuations can become very large, via exchange of energy, number of particles, or volume. Gravity becomes dominant once fluctuations become large. Then, the phenomenon of galaxy clustering may change from linear to nonlinear regime and specific heat may become negative. This change leads to forming bound structures. These large fluctuations are identified with the formation of well defined and sometimes bound clusters of galaxies. Since both number of particles (galaxies) and energy are exchanged across cell boundaries, grand canonocal ensembles can be used for the study.\\

A thermodynamic or macroscopic system undergoes random, rapid and incessant transitions among its microstates. There is a possibility for a cell to find a microstate of higher entropy or lower free energy and ultimately fall into it. Once a cell is trapped in the new microstate, it modifies the thermodynamic functions of its neghbouring cells. A large number of such cells may form nuclei and their modified microstate may propagate throughtout the system. The symmetry of the system may change i.e. a phase transition may occur. The common examples of such phenomena are ice changes to water, water changes to steam, crystals change their structrure and magnetic domains collapse. The large scale observations of the macroscopic systems reveal only the thermodynamic values of the extensive parameters. Interestingly, while in a single phase system the fluctuations are neglegible, they become large and significant in multiphase systems, especially in the neighbourhood of a critical point, at which the fluctuations become large. Near the critical point the fluctuations becmoe so large that there is a high degree of spatial correlation among the molecules of the system making them evident to macroscopic observations as, for example the common phenomenon of crtical opalescence \cite{1998JSP....90..783A, 1993JSP....73..813D, 1995JSP....80...69E, 1995PhRvL..74..208E}. Away from a critical point, the fluctuations are usually small and can be observed only with sophisticated instruments of high resolving power.
Furthermore the theory of fluctuations gives interesting relationships for thermodynamic quantities like, heat capacities. These relationships are used in material science to develop further relations for heat capacities and some other similar quantities.

According to the theory of fluctuations, if x is the fluctuating quantity, then the deviation from its mean value $\bar{x}$ is 
$x-\bar{x}$.
The mean square deviation is a widely used and convenient measure of the magnitude of fluctuation. The mean square deviation is also called the second moment of the distribution.
According to the theory of thermodynamic fluctuations \citep{1960AmJPh..28..684C}, the second order moment is given by

\begin{equation}
 <\Delta X_j\Delta X_k>=\left[\frac{\delta X_k}{\delta F_j}\right]_{F_O...F_{j-1}, F_{j+1}...F_s, X_{s+1}...X_t}
\end{equation}

where
$F_O=\frac{1}{T}$, $F_1=\frac{P}{T}$, $F_2=-\frac{\mu}{T}$,

and

$X_0=U$, $X_1=V$, $X_2=N$\\

We examine the application of fluctuation theory to the cosmological many body problem, in an expanding universe. We consider an infinite system of particles (galaxies) which are clustering gravitationally by their mutual attraction and forming bound clusters. For the cosmological many body system, since the cells can exchange particles (galaxies) as well as energy, grand canonical ensembles are relevant in this work. We imagine a grand canonical ensemble of cells, all of the same volume V, radius R and average number density $\bar {n}$.

Now by suitably arranging the values of constants k and j in equation (1) and using various thermodynamic relations, we calculate the fluctuation moments related to number of particles for single component systems as
\begin{equation}
 <\Delta N^2>=T\left[\frac{\partial N}{\partial \mu}\right]_{T,V}
\end{equation}

Now

\begin{equation}
 \left[\frac{\partial N}{\partial \mu}\right]_{T,V}=\left[\frac{\partial N}{\partial P}\right]_{T,V}\left[\frac{\partial P}{\partial \mu}\right]_{T,V}
\end{equation}

\begin{equation}
 \left[\frac{\partial N}{\partial \mu}\right]_{T,V}=\frac{N}{V}\left[\frac{\partial N}{\partial P}\right]_{T,V}
\end{equation}
where $\frac{\partial P}{\partial \mu}=\frac{N}{V}$

 The equation of state for pressure is \citep{2002ApJ...571..576A}
\begin{equation}
P=\frac{NT}{V}(1-b)
\end{equation}
where $b$ is the clustering parameter for the single component systems \textcolor{red}{\citep{2009AIPC.1150..252M, 2006ApJ...645..940A, 2012MNRAS.421.2629K}}. 
The parameter $b$ measures the influence of gravitational correlation energy W. This parameter is equal to the ratio of gravitational correlation energy W to twice the kinetic energy K of galaxies due to their peculiar velocities in a given volume. It is also related to two point correlation function $\xi (\bar{n},T,r)$.

Differentiating above equation w.r.t. N, we obtain
\begin{equation}
 \left [\frac{\partial P}{\partial N}\right]_{T,V}= \frac{T}{V}\left [1-b-N\frac{\partial b}{\partial N}\right]
\end{equation}

The clustering parameter for single component systems is given by \citep{2002ApJ...571..576A}
\begin{equation}
 b\equiv \frac{-W}{2K} = \frac{\beta \bar{n} T^{-3}}{1+\beta \bar{n} T^{-3}}.
\end{equation}
The parameter $b$ is a measure of the gravitational attraction between particles (galaxies) and is related to the two-point correlation function $\xi (\bar{n},T,r)$ by
\begin{equation}
b=\frac{2\pi Gm^2 \bar{n}}{3T}\int_V \xi (\bar{n},T,r)r{\rm d}r
\end{equation}
The clustering parameter $b$ is esentially the ratio of the gravitational correlation potential energy $W$ to twice the kinetic energy of peculiar velocities $K$ of galaxies on a given scale or volume, $\bar{n} = \bar{N}/V$ is the average number density of galaxies each having mass $m$, and $T$ is the temperature.

Differentiating equation (7) w.r.t. N, we obtain
\begin{equation}
 \frac{\partial b}{\partial N}=\frac{b(1-b)}{N}
\end{equation}
Substituting this relation in equation (6) we obtain
\begin{equation}
 \frac{\partial P}{\partial N}=\frac{T}{V}(1- b)^{2}
\end{equation}

Now using equation (10) in equation (4) we obtain
\begin{equation}
 \left[\frac{\partial N}{\partial \mu}\right]_{T,V}=\frac{N}{T(1-b)^2}
\end{equation}

Finally substituting equation (11) in equation (2) we obtain the second order fluctuation moment in number of particles as

\begin{equation}
 <\Delta N^2>=\frac{N}{(1-b)^{2}}
\end{equation}
In a similar way we can evaluate the higher order number fluctuations as well as the correlated number-energy fluctuations, for single component systems as

\begin{equation}
 <\Delta N^3>=\frac{N}{(1-b)^{4}}(1+2b)
\end{equation}

\begin{equation}
 <\Delta N^4>=\frac{N}{(1-b)^{6}}(1+8b+6b^2)+3(<\Delta N^2>)^2
\end{equation}

\begin{equation}
 <\Delta N\Delta U>=\frac{3}{2}\frac{NT}{(1-b)^2}(1-4b+2b^2)
\end{equation}

The importance of these fluctuation moments stems from the fact that they are used in studying the distribution functions. To determine the shape of the distribution function analytically and to compare it with observations or N body simulations is a primary task in cosmology. Alternatively if it may not be practicable to know the exact shape of the distribution function, its properties can be evaluated by a sequence of the fluctuation moments of the distribution. The moments are also closely related to correlation functions and give important relations for kurtosis and skewness of the distribution of galaxies, which are important because they can be easily related to simulations and observed catalogs \cite{2002ApJ...571..576A}. 

Among all the fluctuation moments, the correlated fluctuations $<\Delta N\Delta U>$ are more important and give interesting results. The value of $<\Delta N>$ can be both positive as well as negative. However, for the present work $<\Delta N>$ is positive, because the clustering parameter $b \ge 0$. The energy fluctuation $<\Delta U>$ can also be either positive or negative. If $<\Delta U>$ is positive then total energy is positive and the kinetic energy is greater than the potential energy. In this case the clustering of particles (galaxies ) is not dominant. On the other hand if $<\Delta U>$ is negative , the total energy is negative. In this case the kinetic energy is less than the potential energy and the clustering of particles (galaxies) is dominant. The critical value of clustering parameter at which this switch occurs is given by

$<\Delta N\Delta U>=0$

Using the above condition in equation (15), gives $b=1/3$ or $T=T_{c}$. This is the value of clustering parameter or critical temperature $T_{c}$ at which specific heat becomes maximum (5/2), then decreases smoothly, becomes zero and finally negative \citep{2012MNRAS.421.2629K}. This suggests that the system of particles (galaxies) becomes essentially relaxed at the maximum value of specific heat and gravitational interactions cause galaxies to cluster. This is also a transition from linear to nonlinear theory of clustering, until the system becomes fully virialized at a value of $b\longrightarrow1$ and specific heat becoming negative (-3/2). It is interesting to note that this is the same value of clustering parameter b or critical temperature $T_{c}$ at which a first order phase transition occurs in the gravitational clustering of galaxies, according to Lee Yang theory (Khan and Malik, 2012, 2013). Reproduction of the earlier results and the exact value of the clustering parameter, validates the application of the fluctuation theory to the cosmological many body problem. The three quite different theories , giving the same resut , lends credibilty to our model and points to some deep symmetry in cosmological evolution process, not fully explored as yet.

\section{Multicomponent Systems}\label{sec3}

The application of fluctuation theory to the gravitational clustering of galaxies has so far been confined to systems having the same mass for all the particles (galaxies) i.e. single component model of galaxies. But this is a simplified model, being far from reality. It is a well known fact that the universe contains clusters having galaxies of different mass spectra. This is also true of groups, superclusters as well as field galaxies. Hence a more accurate model of clustering must take mass spectra into consideration , making a transition from single to multicomponent system of constituent galaxies. This is expected to give more accurate resluts, taking us closer to the real universe. The earlier attempts in this direction have been made by \citet{2009AIPC.1150..252M, 2011Ap&SS.336..447M, 2012MNRAS.421.2629K, 2013Ap&SS.348..211K}, wherein the authors extended the statistical mechanical approach to multicomponent systems, thus analytically deriving the distribution function and the related thermodynamic quantities \cite{2006ApJ...645..940A,2009AIPC.1150..252M,2012MNRAS.421.2629K,2013Ap&SS.348..211K} and \cite{2011Ap&SS.336..447M}. The comparison with the observations has been satisfactory. Thus this is the basic motivation for the present work to apply fluctuation theory to systems having diverse masses. To make a new begining in this direction we use the results of \citet{2009AIPC.1150..252M, 2011Ap&SS.336..447M, 2012MNRAS.421.2629K, 2013Ap&SS.348..211K}. The internal energy for multicomponent systems is given by

\begin{equation}
U=F+TS=\frac{3}{2}NT(1-2b_{m})
\end{equation}
where $b_{m}$ is the clustering parameter for the multicomponent systems\citep{2009AIPC.1150..252M, 2006ApJ...645..940A}

The final expression for internal energy is 

\begin{equation}
U=\frac{3}{2}NT\left[ 1-\frac{2b}{(1+\frac{N_2}{N_1})}\left\{1+\frac{\left(\frac{m_2}{m_1}\right)^3\left(\frac{N_2}{N_1}\right)}{1-b+{\left(\frac{m_2}{m_1}\right)^3 b}}\right\}\right] ,
\end{equation}
where $N_1$ and $N_2$ are the numbers of particles of mass $m_1$ and $m_2$ respectively
so that $N = N_1 + N_2$ is the total number of particles in the whole system.

For incorporating the multicomponent systems into the model, we define two new quatities $\mu$ for $m_2/m_1$
and $\nu$ for $N_2/
N_1$.\\

Simplifying equation (17) we obtain

\begin{equation}
U=\frac{3}{2}T\left[ N_{1}+N_{2}-2bN_{1}-\frac{2b\mu^{3}N_{2}}{1-b+{\mu^{3} b}}\right] ,
\end{equation}

The combination of moments, for multicomponent systems, can be evaluated as follows

\begin{equation}
 <\Delta N_1\Delta U>=\frac{N_1T}{V}\left(\frac{\partial U}{\partial N_{1}}\right) \left(\frac{\partial N_{1}}{\partial P}\right) 
\end{equation}

Differentiating equation (18)
 w.r.t $N_1$ we obtain

\begin{equation}
\begin{split}
 \left(\frac{\partial U}{\partial N_{1}}\right) &= \left[\frac{3}{2}T\right] \\
 & \left[ 1-2b-2N_{1} \left(\frac{\partial b}{\partial N_{1}}\right)-
 2\mu^{3}N_{2}\left(\frac{\partial }{\partial N_{1}}\right)\left(\frac{b}{1-b+{\mu^{3} b}}\right)\right]
 \end{split}
\end{equation}

The pressure for multicomponent systems is given by \citet{2012MNRAS.421.2629K, 2013Ap&SS.348..211K}

\begin{equation}
P=\frac{NT}{V}(1-b_{m})
\end{equation}

After using the expression for clustering parameter $b_{m}$, the final expression for pressure is

\begin{equation}
P=\frac{NT}{V}\left[ 1-\frac{b}{(1+\frac{N_2}{N_1})}\left\{1+\frac{\left(\frac{m_2}{m_1}\right)^3\left(\frac{N_2}{N_1}\right)}{1-b+{\left(\frac{m_2}{m_1}\right)^3 b}}\right\}\right] ,
\end{equation}

Simplifying we get

\begin{equation}
P=\frac{T}{V}\left[ N_{1}+N_{2}-bN_{1}-\frac{b\mu^{3}N_{2}}{1-b+{\mu^{3} b}}\right] ,
\end{equation}
Differentiating w.r.t. $N_1$ we obtain
\begin{equation}
\begin{split}
 \frac{\partial P}{\partial N_{1}} & = \left[\frac{T}{V}\right]\\
 & \left[1-b- N_{1} \left(\frac{\partial b}{\partial N_{1}}\right)
 -\mu^{3}N_{2}\left(\frac{\partial }{\partial N_{1}}\right)\left(\frac{b}{1-b+{\mu^{3} b}}\right)\right]
\end{split}
\end{equation}
Now,
\begin{equation}
 \frac{\partial b}{\partial N_1} = \frac{\partial b}{\partial N}\frac{\partial N}{\partial N_1}
\end{equation}
Since $N = N_1 + N_2$
So, 
\begin{equation}
 \frac{\partial N}{\partial N_1} =1
\end{equation}
Now using equations (9), (25) and (26), we can evaluate the value of the differential 

\begin{equation}
 \frac{\partial }{\partial N_1}\left(\frac{ b}{1-b+{\mu^{3} b}}\right) = \frac{b(1-b)}{N(1-b+{\mu^{3} b)^{2}}}
\end{equation}

Again using equations (25), (26) and (27) in equations (20) and (24), we obtain

\begin{equation}
 \left(\frac{\partial U}{\partial N_{1}}\right) = \frac{3}{2}T\left[ 1-2b-2\frac{b(1-b)}{1+\nu}-2\mu^{3} \nu \frac{b(1-b)}{(1+\nu)(\lambda^{2})}\right]
\end{equation}

\begin{equation}
 \left(\frac{\partial P}{\partial N_{1}}\right) = \frac{T}{V}\left[ 1-b-\frac{b(1-b)}{1+\nu}-\mu^{3} \nu \frac{b(1-b)}{(1+\nu)(\lambda^{2})}\right]
\end{equation}

Now using equations (28) and (29) in equation (19), we obtain

\begin{equation}
\begin{split}
 <\Delta N_1\Delta U> & =\frac{3}{2}N_1T\left[ 1-2b-2\frac{b(1-b)}{1+\nu}-2\mu^{3} \nu \frac{b(1-b)}{(1+\nu)(\lambda^{2})}\right]\\
& \left[\frac{1}{1-b-\frac{b(1-b)}{1+\nu}-\mu^{3} \nu \frac{b(1-b)}{(1+\nu)(\lambda^{2})}}\right]
 \end{split}
\end{equation}

Where $\lambda = 1-b+ \mu^{3} b$

Simplifying further we obtain the correlated number-energy fluctuations for multicomponent systems as

\begin{equation}
\begin{split}
 <\Delta N_1\Delta U> & =\left[\frac{3}{2}N_1T\right]\\
 & \left[ \frac{(1-2b)(\nu +1)\lambda^{2}-2b(1-b)\lambda^{2}-2\mu^{3} \nu b(1-b)}{(1-b)(\nu +1)\lambda^{2}-b(1-b)\lambda^{2}-\mu^{3} \nu b(1-b)}\right]
 \end{split}
\end{equation}

To test the accuracy of our results we can do some checks here. If we equalize the masses of all the galaxies, the above equation, gives the same result as that of equation (15), for single component systems. Alternatively if we ignore the second mass component i.e. $N_2 = 0$, the same result is obtained, thus verifying the accuracy of our model

As expected the correlated fluctuations for a multicomponent system, derived analytically here, give more interesting and informative results as compared to the single component system, described by equation (15). One important result from the above equation is that as the gravitational clustering of galaxies increases, i.e. as the clustering parameter $b\longrightarrow 1$, the fluctuation becomes extremely large. This indicates the formation of bound virialized clusters, indicating that higher the moment, larger the fluctuation and stronger the divergence.

For weak correlations the ensemble average is positive, indicating that a region
of density enhancement typically coincides with a region of positive total energy.
Its perturbed kinetic energy exceeds its perturbed potential energy. For larger correlations the overdense regions typically have negative total energy; underdense regions usually
have positive energy. For multicomponent systems, the critical value at which this switch occurs is given by
 $<\Delta N_1 \Delta U >= 0$. 
Using equation (31) , the result is

\begin{equation}
 \frac{3}{2}N_1T\left[ (1-2b)(\nu +1)\lambda^{2}-2b(1-b)\lambda^{2}-2\mu^{3} \nu b(1-b)\right]=0
\end{equation}
Simplifying further, we finally obtain a quartic equation in clustering parameter b of the form\\
\begin{equation}
Ab^4 + Bb^3 + Cb^2 + Db + E = 0
\end{equation}
where\\
\begin{equation}
A = 2 + 2\mu^6 -4\mu^3\\
\end{equation}
\begin{equation}
B = 4\mu^3\nu-2\nu-2\mu^6
\nu-8-4\mu^6+12\mu^3\\
\end{equation}
\begin{equation}
C = 5\nu+\mu^6\nu-4\mu^3 \nu+\mu^6-10\mu^3+11 \\
\end{equation}
\begin{equation}
D = 2\mu^3-4\nu-6 \\
\end{equation}
\begin{equation}
E = \nu+1
\end{equation}

Here again we can check the accuray of our results for multicomponent systems. Equalizing all the masses or ignoring the second species i.e. ${N_{2}}=0$ and solving the quartic equation (33) for the zeros of b, yields a solution of $b=0.3$, as the only positive real root. This is the same value of clustering parameter b or critical temperature $T_{c}$ at which a transition from positive to negative energy occurs in a single component system of galaxies. Thus the present work for multicomponent systems reproduces all the results of the single component systems, thus confirming our earlier results and the accuracy of equation (33) alongwith all its coefficients from (34) to (38).

\begin{figure*}
\includegraphics[width=0.9\linewidth]{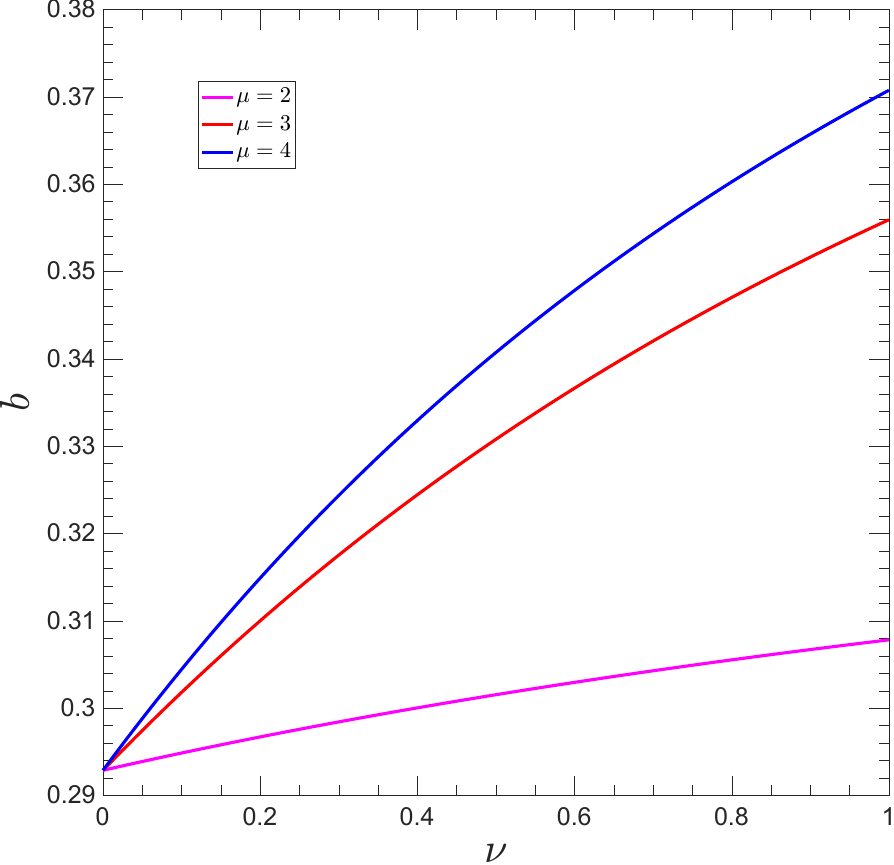}
 \caption{Variation of clustering parameter with the ratio of number densities of different mass components}. 
 \label{fig:nu_b}

\end{figure*}

Figure \ref{fig:nu_b} shows the behaviour of the clustering parameter b as a function of the ratio of the number densities of different mass galaxies, for various mass ratios. Keeping in view the above discussion, it appears that the behaviour of the parameter b will be same, if we plot it as a function of masses, keeping number densities of particles constant. The graph confirms our above discussion in the sense that, as the second mass component is absent, $b\longrightarrow 0.3$, as expected. As defined by Schecter mass function, the galaxy mass function is described by a power law at low masses and an exponential cut-off at high masses. So the number density of more massive galaxies will be always smaller than the number density of less massive galaxies i.e. the parameter $\nu$ will be always smaller than 1 for ($m_2/m_1>1$). Thus the analysis has been limited within the acceptable range of $\nu$ i.e. $n_2/n_1<1$. As the number density ratio increases within this range the clustering parameter shows little increase. However when the ratio of masses $m_2/m_1$ i.e. $\mu$ increases from 2 to 4 to 6, the curves move upwards showing a significant increase. Thus it is clear that the gravitational clustering of galaxies is more sensitive to mass ratios and less sensitive to number densities of galaxies. This means there is little effect of $\nu$ (number density) but significant effect of $\mu$ (mass) of galaxies on the clustering phenomenon. The clustering of galaxies is quicker when mass of individual galaxies increases. They become nuclei for condensation. As the mass ratio of galaxies increases, the transition from positive to negative energy occurs at a higher stage of clustering as compared to a single component system.
Further as the mass ratio of galaxies increases, the value of b at which transition from positive to negative energy occurs, also increases. This is because with increase of mass, correlation potential energy also increases, as most of the total mass of the system is in the more massive galaxies thus dominating the potential energy. Since b measures the correlations between galaxies and is directly proportional to the correlation potential energy, so its value also increases.

\section{Discussion}\label{sec12}
We have investigated the phenomenon of clustering of galaxies in an expanding universe by applying the fluctuation theory. We have evaluated the fluctuation moments for the number of particles (galaxies) as well as the correlated fluctuations for number and energy of particles ( galaxies), clustering under their mutual gravitation. This evaluation has been done, initially for single component system of galaxies. The correlated fluctuations $<\Delta N\Delta U>$ are especially interesting. While as $<\Delta N>$ is positive for this work, $<\Delta U>$, can be both positive as well as negative. For large correlations, the overdense regions typically have negative total energy and underdense regions have usually positive total energy. The critical value at which this switch occurs has been calculated as $b=0.3$, which is equivalent to $T=T_{c}$. It is ineteresting to note that the value of b or $T=T_{c}$, calculated here by fluctuation theory matches with the values calculated in our earlier work by specific heat analysis and Lee Yang theory (Khan and Malik 2012).
The evaluation has been extended to multicomponent systems, consisting of two different mass components. It has been found that the gravitational clustering of galaxies depends uniquely upon the mass ratios and less uniquely on the number densities of galaxies. In line with Schecter mass function, the analysis has been carried out within the acceptable range of $\nu < 1$. It has been observed that there is nominal effect of number density, but considerable effect of $\mu$ (mass) of galaxies on the clustering phenomenon. The increase in mass of galaxies speeds up clustering.
Further as the mass ratio of galaxies increases, the value of b at which transition from positive to negative energy occurs, also increases. This is because with the increase of mass, correlation potential energy also increases, as most of the total mass of the system is in the more massive galaxies thus dominating the potential energy. This can be explained as follows. Since b being the clustering parameter that measures the gravitaional attraction or correlation between galaxies, it is directly proportional to the gravitational correlation potential energy. Therefore as the masses of galaxies increases, the correlation potential energy also increases, leading to a higher value of the clustering parameter at which transition from positive to negative energy occurs. Also more massive galaxies aggregate rapidly and later on the less massive galaxies join them in the phenomenon.

Our study is a begining towards the application of the fluctuation theory to understand the critical phenomena in cosmological many body problem. 
It is expected that our model can be a base for future observational work in cosmology. Especially relating the critical value of cosmological parameter or the critical temperature $T_{c}$, at which clustering of particles becomes dominant, to observations, will be a formidable problem, as it will lead to new insights into the phenomenon of gravitational clustering of galaxies and associated problems.

\section{Data Availability}
We have not used any data in this paper. All the results generated in this paper are theoretical.
\bibliography{sn-bibliography}

\end{document}